\documentclass[showpacs,preprintnumbers,amsmath,amssymb,superscriptaddress]{revtex4}
\usepackage{amsmath}
\usepackage{mathrsfs}
\usepackage{amssymb}
\usepackage{revsymb}
\usepackage{graphicx} 
\begin{document}

\title{SU(1,1) symmetry of multimode squeezed states}

\author{Z Shaterzadeh-Yazdi, P S Turner and B C Sanders}
\address{Institute for Quantum Information Science, University of Calgary, 2500 University Drive NW, Calgary, Alberta, Canada, T2N 1N4}
\email{zyazdi@qis.ucalgary.ca}, \email{pturner@qis.ucalgary.ca}

\begin{abstract}
We show that a class of multimode optical transformations that employ linear optics plus
two-mode squeezing can be expressed as SU(1,1) operators.
These operations are relevant to state-of-the-art continuous variable quantum information
experiments including quantum state sharing, quantum teleportation, and multipartite entangled states.
Using this SU(1,1) description of these transformations, we obtain a new basis for
such transformations that lies in a useful representation of this group and lies 
outside the often-used restriction to Gaussian states.
We analyze this basis, show its application to a class of transformations, and discuss
its extension to more general quantum optical networks.
\end{abstract}
\vspace{2pc}

\maketitle

\section{Introduction}

The Heisenberg uncertainty principle ensures that light is noisy at a quantum level.
Mathematically, this noise arises because of non-commutativity of the sine and cosine 
quadratures of light. In the vacuum state, the noise in both quadratures are equal, but 
the noise can be squeezed -- i.e. the noise in one quadrature can be reduced at the 
expense of increasing noise in the complementary quadrature -- by propagation through
nonlinear optical devices such as crystals and gases. In quantum information applications,
squeezers, which shift fluctuations from one quadrature to another~\cite{loudon}, provide an
entanglement resource through correlated noise in two modes with these correlations
stronger than anything possible in a classical description of light~\cite{caves, schumaker}. 

The importance of squeezed light in quantum information is evident in its role in some of the most important experiments in quantum information science: teleportation~\cite{furusawa}, entanglement swapping~\cite{glock}, tests of local realism and Einstein-Podolsky-Rosen (EPR) paradox~\cite{ou92a}, and quantum state sharing~\cite{lance2003,tyc2002}. These continuous variable quantum information experiments involve more than two modes, and entanglement between more than two modes is inherent in these experiments.

Here we analyze multimode squeezed light.
Specifically, we are interested in the case of two-mode squeezed light
(which could be generated by a two-mode squeezer or, alternatively, by two
single-mode squeezers in opposite phase with their outputs mixed at a beam
splitter to yield two-mode squeezed light~\cite{kim96}), with the squeezed field
distributed over multiple modes via linear optics to share the squeezed light between
modes. Mathematically, the transformation is described by a matrix, which transforms
input state vectors into output state vectors. The choice of basis can simplify the mathematical 
description as well as the calculations themselves. For example, if the inputs are Gaussian
states, it is convenient to use a Gaussian basis, and transformations can be fully described
in terms of the multivariate vector of means and the multivariate covariance matrix~\cite{plenio}.
For general input states, the Fock number state basis is a useful basis to write general transformations;
two alternative representations that transform simply under squeezing and linear optics are 
provided by the Wigner function~\cite{kimble, simon87} and by the position representation~\cite{tyc2002}. The problem with these approaches for general input states is that the size of the matrix grows linearly in the number of modes.

On the other hand, the use of group theoretical concepts in quantum optics is becoming more and more prevalent~\cite{yurke, cervero, wunsche00, puri94}. Understanding the symmetry of a quantum system and identifying its group properties allow one to use a large class of mathematical tools to simplify problems~\cite{bargmann}. Here, we show that an alternative basis can be obtained,
which relies on determining the symmetries of the network
transformation and using the powerful machinery of Lie group theory,
including the Wigner-Eckart theorem~\cite{wigner59}.
In our basis, the mathematical description has constant size independent of the number of modes.
Our approach builds upon the SU(1,1) symmetry inherent in combining squeezing 
with linear optics in two modes.

In the absence of squeezing, $n$-mode linear optical transformations are elements 
of the compact special unitary group SU(n)~\cite{aniello05}; with squeezers included in the network, 
optical network transformations are
given by elements of the symplectic group~Sp($2n,\mathbb{R}$)~\cite{plenio, steve02}. The fact that these transformations
are members of the symplectic group guarantees the simplicity of describing Gaussian state 
transformations using just covariance matrices and transformations of means~\cite{simon88}, but this 
simplicity does not carry over to general non-Gaussian states.

Here, we show that a broad class of quantum optical networks involving two-mode squeezing
and linear optics of arbitrarily many modes, can be greatly simplified by exploiting a
SU(1,1) $\cong$Sp($2,\mathbb{R}$)$\subset$ Sp($2n,\mathbb{R}$) symmetry in such systems.
This symmetry allows us to describe the transformation with fixed size independent of
the number $n$ of modes.
All examples mentioned above -- quantum teleportation, entanglement swapping, state
sharing, and tests of local realism -- have transformations that are members of this class.
We exploit this SU(1,1) symmetry by finding a basis that reduces the $n$-mode Fock states
into irreducible representations (irreps) of this group.

In addition to finding the SU(1,1) symmetry for a broad class of interferometers comprising
linear optics and two-mode squeezing, and finding a convenient basis in which the
transformations have fixed size independent of the number~$n$ of modes, our work points
to new directions in studies of more complex interferometers involving more squeezers.
The Bloch-Messiah theorem applied to quantum optical networks~\cite{rowe, braunstein} is a powerful tool to reduce such interferometers to all linear optical transformations, followed
by single-mode squeezers, followed by all linear optical transformations. 
The approach we establish here may provide an alternative to decomposing such
interferometers by concatenating interferometers into a larger whole.
We discuss this possible future direction for research in the Conclusions.

\section{Background: Two-mode squeezing and the SU(1,1) Lie group}\label{twomodesection}

Two-mode squeezed states can be generated either by entangling two independent single-mode squeezed states via a 50:50 beamsplitter, or by employing the non-degenerate operation of a non-linear medium in the presence of two incoming modes~\cite{kim}.  
The unitary operator describing two-mode squeezing is  
\begin{equation}\label{squeezer}
\hat S_{ab}(\eta)=\exp\left[-\text{i}(\eta\hat a\hat b+\eta^*\hat a^{\dagger}\hat b^{\dagger})/2\right],
\end{equation}
with $\hat a$, $\hat b$ the annihilation operators of the incoming modes
and $\eta \in \mathbb{C}$ the squeezing parameter. This operator gives a unitary representation of the SU(1,1) Lie group on the Hilbert space of two modes. As such, it is generated by a $\mathfrak{su}(1,1)$ Lie algebra given by
\begin{equation}
\hat K_+ = \hat a^\dag \hat b^\dag, \qquad
\hat K_- = \hat a \hat b, \qquad
\hat K_0 = \frac{1}{2}\left(\hat a^\dag \hat a + \hat b \hat b^\dag \right),
\end{equation}
which satisfy the commutation relations
\begin{equation}
\left[ \hat{K}_0,\hat{K}_\pm\right] = \pm \hat{K}_\pm, \qquad
\left[ \hat{K}_-,\hat{K}_+\right] = 2\hat{K}_0. 
\end{equation}
The SU(1,1) Casimir invariant is 
\begin{equation}
\hat K^2=\hat K_0^2-\frac{1}{2}(\hat K_+\hat K_-+\hat K_-\hat K_+) = \frac{1}{4}\left[(\hat a^\dag \hat a - \hat b^\dag \hat b)^2-\hat{\openone}\right].
\end{equation}

Eigenvalues of $\hat K^2$ are used to label the irreps of SU(1,1),
and eigenvalues of $\hat K_0$ provide an index for a basis of each irrep.
Denoting such an orthonormal basis by $\{\vert k,\mu\rangle\}$, we have the following SU(1,1) action
\begin{eqnarray}
\hat K_\pm \vert k,\mu\rangle&=&\sqrt{(\mu\pm k)(\mu\mp k\pm1)} \vert k,\mu\pm1\rangle,\\
\hat K_0 \vert k,\mu\rangle&=&\mu\vert k,\mu\rangle,\label{cartan}\\
\hat K^2 \vert k,\mu\rangle&=&k(k-1)\vert k,\mu\rangle.\label{casimir}
\end{eqnarray}
Due to the non-compactness of this group, all unitary irreps are infinite dimensional. There are in fact several different series of irreducible representations of SU(1,1) distinguished by the domains of these eigenvalues~\cite{bargmann}. For now, we are only interested in the usual positive discrete series of irreps where $k$ is a non-negative half integer and $\mu$ takes values $k+m$ for $m=0,1,2,\cdots$ carried by a Hilbert space denoted by $\mathscr{D}^+_k$.  

If we label the number of excitations for modes $a$ and $b$ by $n_a$ and $n_b$, respectively, then $k$ satisfies
\begin{equation}
\label{eq:ksatisfies}
	k = \frac{\vert n_a-n_b\vert+1}{2}.
\end{equation}
As the relabelling of modes $a\leftrightarrow b$ is physically equivalent to the original labelling, we consider the cases $\pm (n_a-n_b)$ to be equivalent irreps (hence the absolute value above).  Thus we say that each irrep $k > 0$ occurs twice, and the Hilbert space decomposes as
\begin{equation}
\mathscr{H}_a\otimes\mathscr{H}_b
 = \mathscr{D}^+_\frac{1}{2} \oplus 2 \mathscr{D}^+_1 \oplus 2 \mathscr{D}^+_\frac{3}{2} \oplus \cdots
\end{equation}
Assuming that mode $a$ has more photons than mode $b$, an arbitrary two-mode state $ \vert n_a ,n_b\rangle$, is equivalent to SU(1,1) weight states $\vert k,\mu\rangle$ with
\begin{equation}\label{twomodelabels}
k = \frac{ n_a-n_b+1}{2}, \qquad \mu = \frac{n_a+n_b+1}{2}.
\end{equation}
As the irrep label $k$ is proportional to the photon number difference,
and irrep spaces are invariant under SU(1,1),
the photon number difference is conserved for two-mode squeezers.
In other words, the photon number difference in the two modes entering these optical elements equals the photon number difference leaving them.

Discovering an appropriate realization of a symmetry group enables a clear understanding of a system via the mathematical properties that are already known for these groups.
We can exploit representation-theoretic machinery, such as selection rules and branching rules to facilitate calculations~\cite{bargmann, biedenharn, rowerepka}.
Output states are then characterized by generalized coherent states that are known for groups such as SU(1,1)~\cite{perelomov72}. 

Often, understanding the symmetry of a system also brings with it much needed physical insight as systems become more complicated.  However, we will limit the scope of this paper to identifying that symmetry and constructing representations for a class of linear optical networks plus squeezers that occur in several important optical quantum information protocols as described in the next section.

\section{Motivation: Three-mode squeezing and the SU(1,1) Lie group}

\begin{figure}
\centering
\includegraphics[width=15cm, height=12cm]{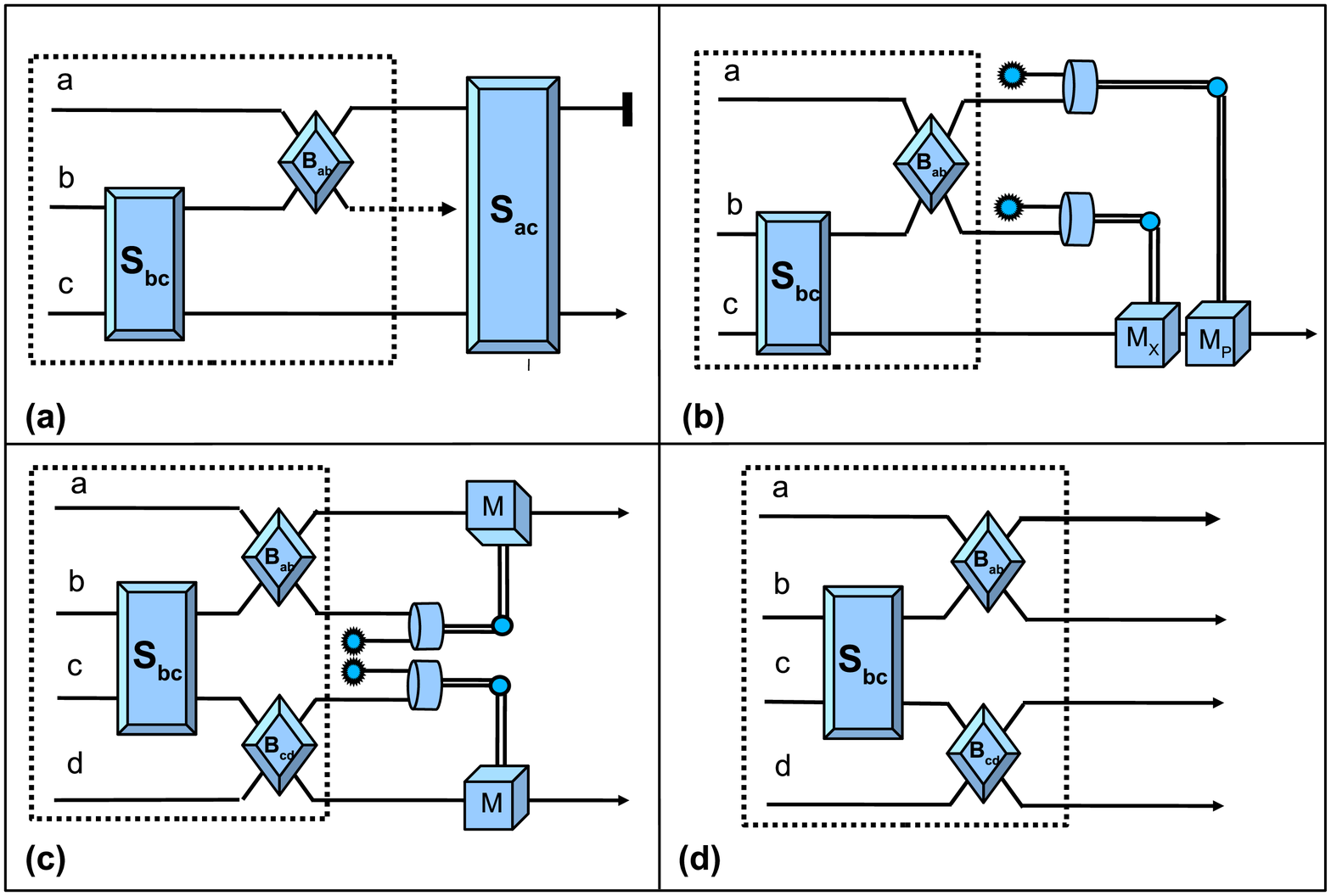}
\caption {(Colour online) The dashed rectangles show that how three-mode or four-mode squeezed states, which are a key source of entanglement, can be generated by distributing a two-mode squeezed state over other modes. Such states are a required ingredient for various quantum information tasks such as: (a)~quantum state sharing~\cite{rowe};
(b)~quantum teleportation~\cite{kimble}; 
(c)~entanglement swapping~\cite{glock}; and
(d) tests of local realism and Einstein-Podolosky-Rosen paradox~\cite{ou92a}.}\label{figure1a_1d}
\end{figure}

\begin{figure}
\centering
\includegraphics[width=7cm, height=5cm]{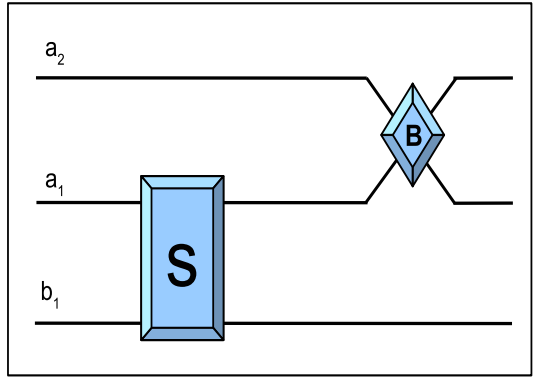}
\caption {(Colour online) The `primitive' three-mode optical network in which tripartite entangled states are produced. Such states are the first step towards multipartite entangled states and are applied in various optical quantum schemes some of which are shown in Fig.~\ref{figure1a_1d}.}
\label{figure2}
\end{figure}

As mentioned in the Introduction,
the motivation for this research stems from recent optical quantum information experiments.
In Fig.~\ref{figure1a_1d} we have included simple schematics of these experiments.
As is indicated in the figure, all these systems incorporate a two-mode squeezer
with one (or both) of the squeezer's output modes mixed with a third (and fourth) mode on a 50:50 beam splitter, thereby distributing entanglement across three (or four) modes in the network; see Fig.~\ref{figure2}.

Two-mode squeezing action is given in Eq.~(\ref{squeezer});
the beamsplitter action on modes $a$ and $b$ is given by
\begin{equation}\label{splitter}
	\hat B_{ab}(\theta,\phi) 
		= \exp\left[\theta\left(\hat a^{\dagger}\hat b\text{e}^{\text{i}\phi}
			-\hat a\hat b^{\dagger}\text{e}^{-\text{i}\phi}\right)/2\right]\,.
\end{equation}
For the given values of the arguments, the transformation performed by the three-mode component of Fig.~\ref{figure2} is~\cite{kim}
\begin{equation}
	\hat B^+_{a_2a_1}\hat S_{a_1b_1}(2\text{i}\eta)
		=\hat S_{a_2a_1b_1}(\sqrt{2}\text{i}\eta)\hat B^+_{a_2a_1}
\end{equation}
for
\begin{equation}
	\hat B^\pm_{ab}\equiv\hat B_{ab}\left(\frac{\pi}{2},\pm\pi\right)
\end{equation}
and
\begin{equation}
	\hat S_{a_2a_1b_1}\left(\sqrt{2}\text{i}\eta\right)
		= \exp{\left[\frac{-\text{i}\eta}{\sqrt{2}}\left(\hat a_1\hat b_1-\hat a_2\hat {b_1}\right)
			+\frac{\text{i}\eta^*}{\sqrt {2}}\left(\hat a_1^{\dagger}\hat b_1^{\dagger}
				-\hat a_2^{\dagger}\hat b_1^{\dagger}\right)\right]}.
\end{equation}
It is not difficult to check that the generators of this transformation satisfy a $\mathfrak{su}(1,1)$ algebra with ladder operator
\begin{equation}
	\hat K_+
		=\frac{1}{\sqrt{2}}\left(\hat a_1^{\dagger}\hat b_1^{\dagger}
			-\hat a_2^{\dagger}\hat b_1^{\dagger}\right).
\end{equation}
The specifics of this realization have discussed in our preliminary investigation~\cite{spie}.
In the next section we generalize this result to arbitrarily many modes.

\section{Results: multimode squeezing and the SU(1,1) Lie group}

In this section, we show that the three-mode entanglement distributing component given 
in Fig.~\ref{figure2} can be extended to arbitrarily many modes while still being generated by a 
$\mathfrak{su}(1,1)$ algebra.  We then analyze the representations of this algebra on the multimode Fock space, giving the new basis of SU(1,1) weight states.

\begin{figure}
\centering
\includegraphics[width=15cm, height=12cm]{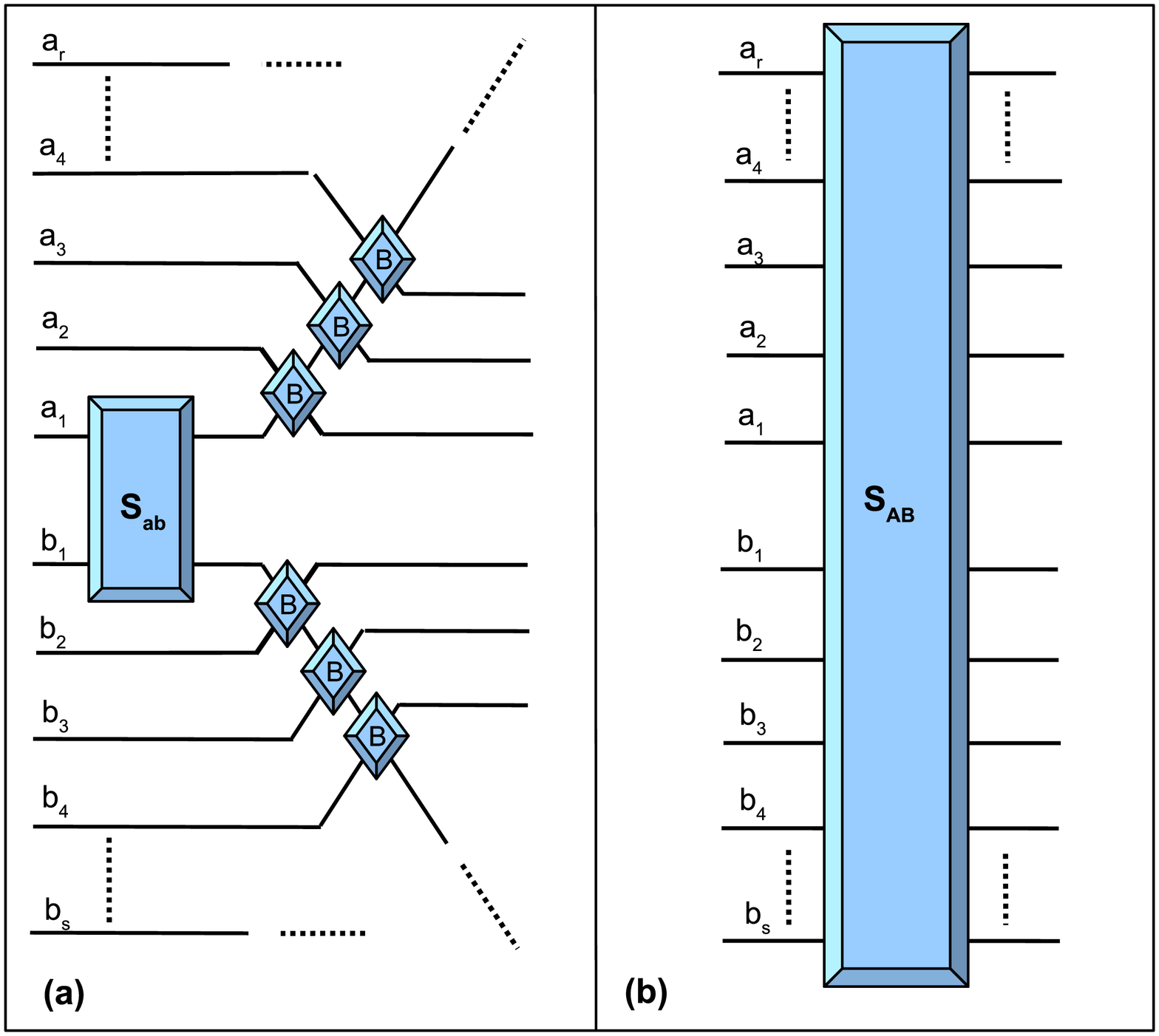}
\caption [l]{(Colour online) Equivalence between a typical multimode quantum optical network and a pseudo-two-mode squeezer. (a) A multimode optical network comprising a two-mode
squeezer ($S_{ab}$) and several 50:50 beam splitters. (b) A pseudo-two-mode squeezer ($S_{AB}$).}
\label{figure3a_3b}
\end{figure}

We are considering an optical network (see Fig.~\ref{figure3a_3b}) that comprises one two-mode squeezer in which one output state is mixed via beam splitters between $r$ modes, created by $\{\hat a_l^\dag\}_{l=1}^r$, and the other mixed with $s$ modes, created by $\{\hat b_l^\dag\}_{l=1}^s$.  
For simplicity we consider only 50:50 beam splitters with beam splitters for the upper~$r$ modes having a phase $\phi=\pi$ and those for the lower~$s$ modes having $\phi=-\pi$. 
Based on the Baker-Campbell-Hausdorff formula \footnote{The Baker-Campbell-Hausdorff formula is $
e^{\text{i}\lambda \hat{A}} \hat{B} e^{-\text{i}\lambda \hat{A}} = \hat{B} + {(\text{i}\lambda)}[\hat{A},\hat{B}] + \frac{(\text{i}\lambda)^2}{2!} [\hat{A},[\hat{A},\hat{B}]] +\frac{(\text{i}\lambda)^3}{3!} [\hat{A},[\hat{A},[\hat{A},\hat{B}]]]+\cdots$}, the resulting transformation performed by the multimode network is
\begin{eqnarray}
\label{BScommute}
	\hat B^+_{a_ra_{r-1}}\cdots\hat B^+_{a_2a_1}
	\hat B^-_{b_sb_{s-1}}\cdots\hat B^-_{b_2b_1}
	\hat S_{a_1 b_1}(\eta)
		=\hat S_{A_r B_s}(\eta)
	\hat B^+_{a_ra_{r-1}}\cdots\hat B^+_{a_2a_1}
	\hat B^-_{b_sb_{s-1}}\cdots\hat B^-_{b_2b_1},
\end{eqnarray}
with $\hat S_{A_rB_s}$ being a multimode squeezing operator (see Fig.~\ref{figure3a_3b}). This operator can be viewed as a Bogoliubov-transformed two-mode squeezer and is given by
\begin{equation}
\hat S_{A_rB_s}=\exp\left[-\text{i}(\eta\hat A_r\hat B_s+\eta^*\hat A_r^{\dagger}\hat
B_s^{\dagger})/2\right],
\end{equation}
where $\hat A_r$ and $\hat B_s$ are the generalized boson (\emph{pseudo-boson}) operators 
\begin{equation}
\label{eq:ArBs}
\hat A_r=\sum_{l=1}^{r-1}\frac{(-1)^{l-1}\hat a_l}{\sqrt{2^l}}+\frac{(-1)^{r-1}\hat a_r}{\sqrt{2^{r-1}}},\qquad
\hat B_s=\sum_{l=1}^{s-1}\frac{(-1)^{l-1}\hat b_l}{\sqrt{2^l}}+\frac{(-1)^{s-1}\hat
b_s}{\sqrt{2^{s-1}}}
\end{equation}
given in terms of the original optical modes $a_l$ and $b_l$ for $r,s\geq 2$. Of course, these pseudo-operators satisfy the canonical bosonic commutation relations
\begin{equation}
[\hat A_r,\hat A_r^{\dagger}]=\hat {\openone}, \qquad[\hat B_s, \hat B_s^{\dagger}]=\hat {\openone}.
\end{equation}

The Hamiltonian generating $\hat S_{A_r B_s}$ is a linear combination of the two operators $\hat A_r\hat B_s$ and $\hat A_r^{\dagger}\hat B_s^{\dagger}$. 
These together with their commutators are closed under the commutation relations of a $\mathfrak{su}(1,1)$ algebra and thus provide a pseudo-two-boson realization of $\mathfrak{su}(1,1)$
\begin{equation}
\hat K_-=\hat A_r\hat B_s, \qquad 
\hat K_+= \hat A_r^{\dagger}\hat B_s^{\dagger},\qquad
\hat K_0=\frac{1}{2}(\hat A_r^{\dagger}\hat A_r+\hat B_s^{\dagger}\hat B_s+ \hat {\openone}).
\end{equation}
The Casimir operator for this realization is
\begin{equation}
\hat K^2=\frac{1}{4}\left[(\hat A_r^{\dagger}\hat A_r-\hat B_s^{\dagger}\hat B_s)^2-\hat {\openone}\right]
\end{equation}
with $\hat A_r^{\dagger}\hat A_r$ and $\hat B_s^{\dagger}\hat B_s$ the pseudo-number operators,
$\hat N_A$ and $\hat N_B$, respectively.
If we define pseudo-number states such that
\begin{equation}
\hat N_A\vert n_A\}=n_A\vert n_A\},\qquad \hat N_B\vert n_B\}=n_B\vert n_B\},
\end{equation}
then the entire situation is formally equivalent to the two-mode realization given in the Section~\ref{twomodesection}.  From Eq.~(\ref{twomodelabels}), we can read off the labels for a basis of SU(1,1) weight states $\vert k,\mu\rangle$ in terms of the pseudo-number states $\vert n_A n_B\}$:
\begin{eqnarray}
k&=&\frac{1}{2}\left(\vert n_A-n_B\vert+1\right), \qquad\qquad \mu=k+m=\frac{1}{2}(n_A+n_B+1),\\
k&=&\frac{1}{2}, 1, \frac{3}{2}, \cdots, \qquad\qquad\qquad\;\;\; m=0, 1, 2,\cdots,\\
n_A&=&n_{a_1}+n_{a_2}+\cdots+n_{a_r},\qquad n_B=n_{b_1}+n_{b_2}+\cdots+n_{b_s}.
\end{eqnarray}
Therefore, the difference between the total photon number for modes $a_1$ to $a_r$ and the total photon number for modes $b_1$ to $b_s$ is conserved.

All that remains in order to specify the new basis is to give the pseudo-number states in terms of the original modes.  For $\vert n_A\}$, and analogously for $\vert n_B\}$, we obtain
\begin{equation}\label{pseudocoeff}
\vert n_A\}=\frac{\left( A_r^\dag \right)^{n_A}}{\sqrt{n_A!}} \vert 0\} = \sum_{\mathbf{n}} C_{\mathbf{n}}^{n_{A}} \vert \mathbf{n} \rangle,
\end{equation}
where $\mathbf{n}=(n_{a_1},n_{a_2},\cdots,n_{a_r})$ is a partition of $n_A$:
\begin{equation}
\sum_{l=1}^r n_{a_l} = n_A, \qquad 0 \leq n_{a_l} \in \mathbb{Z}\,,
\end{equation}
and $\vert\mathbf{n}\rangle =\vert n_{a_1},n_{a_2},\cdots,n_{a_r}\rangle$ is a multimode number state.  From Eqs.~(\ref{eq:ArBs}) and~(\ref{pseudocoeff}) we have
\begin{equation}
C_{\mathbf{n}}^{n_{A}}
=\left(\frac{n_A!}{n_{a_1}!n_{a_2}!\cdots n_{a_r}!}\right)^{1/2}
\left[ \prod_{l=1}^{r-1} \left(\frac{(-1)^{l-1}}{\sqrt{2^l}}\right)^{n_{a_l}} \right] \left( \frac{(-1)^{r-1}}{\sqrt{2^{r-1}}} \right)^{n_{a_r}}.
\end{equation}
It is not difficult to verify that these states are orthonormal, $\{n_A\vert m_A\}=\delta_{n_A m_A}$.

Take for example the experimentally significant three-mode case depicted in Fig.~\ref{figure2}.  In this case, the pseudo-number states, in terms of the original modes $a_1$ and $a_2$, are given by
\begin{equation}
	\vert n_A \} = \frac{1}{\sqrt{2^{n_A}}} \sum_{n_{a_1}
		=0}^{n_A} (-1)^{N_A-n_{a_1}} {n_A\choose{n_{a_1}}}^{1/2}
		\vert n_{a_1} n_{a_2}\rangle, \qquad n_A=n_{a_1}+n_{a_2},
\end{equation}
which is normalized because
\begin{equation}
	\frac{1}{2^{n_A}} \sum_{n_1=0}^{n_A}{n_A\choose{n_{a_1}}}=1.
\end{equation}
Assuming that the number of photons in modes $a_1$ and $a_2$ exceeds the photon number in mode $b_1$,
the $\mu^{\rm th}$ three-mode SU(1,1) weight state in irrep $k$, in terms of these three photon number states, is 
\begin{equation}
\vert k,\mu\rangle=\frac{1}{\sqrt{2^{k+\mu-1}}} \sum_{n_{a_1}=0}^{k+\mu-1} (-1)^{n_{a_1}}{k+\mu-1\choose{n_{a_1}}}^{1/2}\vert k+\mu-n_{a_1}-1, n_{a_1}, \mu-k\rangle.
\end{equation}
However, by this assumption we can only get half of the irreps. To get the other half, one needs to exchange $k+\mu-1$ and $\mu-k$ in the above expression.  

\section{Discussion}

\begin{figure}
\centering
\includegraphics[width=7cm, height=10cm]{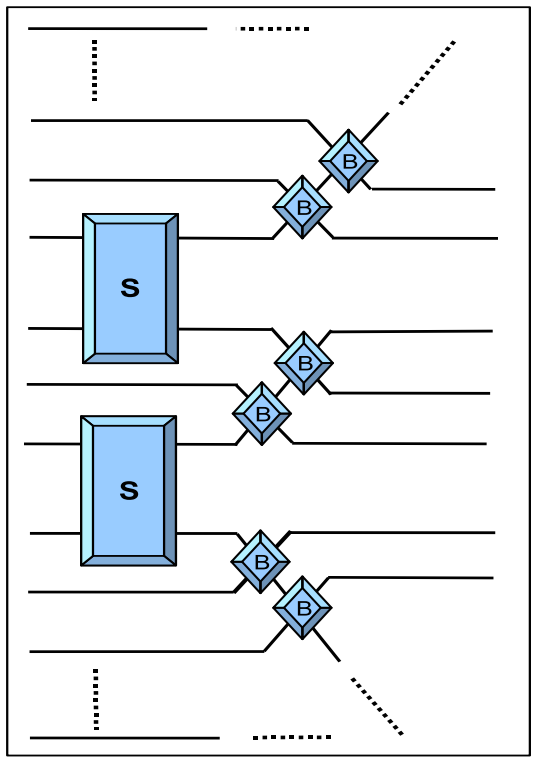}
\caption {(Colour online) A multimode network with more than one two-mode squeezer. Such a network can be built by concatenating different parellel sections each being of the type of multimode network considered earlier. The known SU(1,1) Clebsch-Gordan and Racah coefficients can be used to couple the corresponding representations and to obtain the resultant states.}
\label{figure4}
\end{figure}

This SU(1,1) basis enables us to use a variety of mathematical properties that have already been established for this group, thereby significantly facilitating calculations. 
For instance, the SU(1,1) Clebsch-Gordan coefficients~\cite{biedenharn, wang, gerry04} are known for coupling different representations and obtaining the resultant states. These coefficients are very handy if one wants to concatenate some of these typical multimode networks `in parallel' (see Fig.~\ref{figure4}). Furthermore, if the concatenation takes place between more than two such optical networks, then the Racah coefficients~\cite{fano} give the transformations between different orderings of the couplings. The Wigner-Eckart theorem~\cite{wigner59} makes the calculation of matrix elements of a tensor operator, or any operator generated from the elements of the algebra ( a linear combination thereof, such as the squeezing operator) in this SU(1,1) basis much simpler than a direct approach. Generalized coherent states~\cite{perelomov72}, orthogonality, and asymptotic behaviour of the matrix elements~\cite{bargmann} are some other examples of the well known properties for this 
 group.

The significance of our result is that it allows for arbitrary input states, in contrast to existing methods which usually rely upon Gaussian inputs.  This technique can therefore be used to characterize a larger class of output states.  Moreover, because the algebraic structure is independent of the number of modes, complicated multimode linear quantum optical networks with two-mode squeezing,
that are of this form, can be dealt with much more easily than with, for example, Wigner functions, in which one needs to calculate $2n$ integrals, or covariance matrices for which $4n^2$ matrix elements must be calculated, where $n$ is the number of modes.

These advantages come with a cost, namely a multiplicity of SU(1,1) irreps that grows as the number of modes.  For $r+s$ modes there are always only two pseudo-boson operators, $\hat {A}_r$ and $\hat {B}_s$, leaving $r+s-2$ `directions' in Fock space.
Finding $r+s-2$ orthogonal pseudo-boson operators and using them to label multiplicities would be tedious but not impossible for large $r$ or $s$:
this is an open question for further research.  Also, we have limited our attention to linear quantum 
optical networks with two-mode squeezing with the specific sequence of beamsplitters 
given by Eq.~(\ref{BScommute}) --- we considered this form since it is a direct extension of the networks in existing experiments.  This SU(1,1) realization will occur for any Bogoliubov transformation that mixes each output port of the two-mode squeezer separately; that is, any Bogoliubov transformation that does not mix $a$ modes and $b$ modes (see Fig.~\ref{figure3a_3b}).  When these modes are mixed `in series' by some multiport optical element a far more complicated situation arises, as 
exemplified in the next section.

\section{More complicated scenarios}

Consider the network arising from the reconstruction protocol of the quantum secret sharing experiment in Fig.~\ref{figure1a_1d}.  Here,  there is an extra two-mode squeezer operating on one mode from the upper set and one mode from the lower set of the three-mode squeezing scheme given in Fig.~\ref{figure2}. Consequently, we obtain the identity
\begin{equation}
\hat S_{a_2b_1}(\eta')\hat S_{a_2a_1b_1}(\sqrt{2}\text{i}\eta)= \hat {\mathcal{S}}_{a_2a_1b_1}(\eta, \eta')\hat S_{a_2b_1}(\eta'),
\end{equation}
where $\hat {\mathcal{S}}_{a_2a_1b_1}(\eta, \eta')$  is the exponential of a linear combination of twelve terms.  However, by fixing two of the parameters,
\begin{equation}
\phi'=\frac{\pi}{2}\Rightarrow\eta'=s'\exp(\text{i}\phi')=\text{i} s',
\eta=\eta*\Rightarrow\phi=2m\pi
\end{equation}
for $m=0, 1, 2, \cdots$, we obtain
\begin{eqnarray}
\hat {\mathcal{S}}_{a_2a_1b_1}(\eta, \eta')=\exp\left[\frac{\eta}{\sqrt{2}}\cosh\left(\frac{s'}{2}\right)(\hat{a}_1\hat{b}_1-\hat{a}^{\dagger}_1\hat{b}^{\dagger}_1)+\frac{\eta}{\sqrt{2}}\sinh\left(\frac{s'}{2}\right)(\hat{a}^{\dagger}_2\hat{a}_1-\hat{a}_2 \hat{a}^{\dagger}_1)-\frac{\eta}{\sqrt{2}}(\hat{a}_2\hat{b}_1-\hat{a}^{\dagger}_2\hat{b}^{\dagger}_1)\right].
\end{eqnarray}
This operator is generated by
\begin{equation}
\hat K_x=-\text{i}(\hat{a}_1\hat{b}_1-\hat{a}^{\dagger}_1\hat{b}^{\dagger}_1),\qquad
\hat K_y=-\text{i}(\hat{a}_2\hat{b}_1-\hat{a}^{\dagger}_2\hat{b}^{\dagger}_1),\qquad
\hat K_z=\text{i}(\hat{a}^{\dagger}_2\hat{a}_1-\hat{a}_2 \hat{a}^{\dagger}_1),
\end{equation}
which satisfy the commutation relations of $\mathfrak{su}(1,1)$. 
A unitarily equivalent form of these operators has been found previously 
by Sebawe Abdalla et al.~\cite{abdalla}. One can unitarily transform these operators using 
\begin{equation}
\hat U=\exp\left [\text{i}\frac{\pi}{4}(\hat{a}^{\dagger}_2\hat{a}_1+\hat{a}_2 \hat{a}^{\dagger}_1)\right],
\end{equation}
in order to diagonalize $\hat K_z$ in the Fock basis:
\begin{equation}
\hat {\tilde{K}}_z=\hat U\hat K_z\hat U^{\dagger}=\hat{a}^{\dagger}_2\hat{a}_2- \hat{a}^{\dagger}_1\hat{a}_1.
\end{equation}
However, this transformation does not simplify the form of the other operators;
in particular, the Casimir operator
\begin{eqnarray}
\hat {\tilde K}^2=\hat U\hat K^2\hat U^{\dagger}=(\hat{a}^{\dagger}_2\hat{a}_2- \hat{a}^{\dagger}_1\hat{a}_1)^2
	-2(\hat{a}^{\dagger}_2\hat{a}_2+\hat{a}^{\dagger}_1\hat{a}_1+\hat{\openone}) 
		\hat{b}^{\dagger}_1\hat{b}_1
			-(\hat{a}^{\dagger}_2\hat{a}_2+\hat{a}^{\dagger}_1\hat{a}_1+2\hat{\openone})
				+2\text{i}(\hat{a}^{\dagger}_2\hat{a}^{\dagger}_1\hat{b}^{\dagger2}_1
					-\hat{a}_2\hat{a}_1 \hat{b}^2_1).
\end{eqnarray}

Identifying the appropriate representations of this realization of $\mathfrak{su}(1,1)$ carried by the Fock space of three modes is quite challenging.  For one, the lowering operator $\hat{K}_-$ is a linear combination of both lowering \emph{and} raising operators.  This means that the lowest weight $\mu=k$ states are not Fock states but superpositions of Fock states, 
a situation that does not occur in the standard one- and two-mode realizations. 
More importantly, $\hat K_z$ can have both positive and negative eigenvalues in the Fock basis, which, along with the complicated structure of $\hat K^2$, suggests that we have left the realm of discrete SU(1,1) irreps and must consider the less intuitive continuous irreps that do not have extremal weights.  The problem of which classes of SU(1,1) irreps are supported on multimode Fock spaces is an interesting one deserving further study.  Understanding these representations would give us a powerful analytical tool since it would give a complete symmetry adapted basis for these and ultimately more complicated optical networks, which would aid in the investigation of squeezing and entanglement therein.

\section{Conclusions}
\label{sect:conclusion}

We have established a novel basis lying in the discrete representation of SU(1,1). This basis is adapted to physical problems of multimode squeezing. Such problems occur in various experimental set-ups for optical quantum information schemes involving a two-mode squeezer and several passive optical elements. This technique facilitates the calculation of output states without restricting to `standard' input states like Gaussians.

We also showed that some interesting mathematical problems arise in more complicated multimode schemes. These problems seem inherent to linear quantum optical networks with squeezers and
with more than two mixed modes; e.g. it also arises in a four-mode realization of squeezing studied by Bartlett et al.~\cite{bartlett}. This interesting complication opens questions about whether,
by concatenating optical networks, each with just one two-mode squeezer,
the description remains within a discrete series representation of SU(1,1). Based on the evidence discussed in previous section, it seems unlikely. Such problems are important in considering practical quantum information tasks, in studying squeezing as a resource and in similar problems; our work 
reported here is an important step in this direction.

\section*{ACKNOWLEDGMENT}

We appreciate valuable, insightful discussions with Stephen Bartlett in the early stages of this research. This work has been supported by iCORE, CIFAR, NSERC, AIF and MITACS.

\bibliography{iopbiblio_02.bib}
\bibliographystyle{iopart-num.bst}

\end{document}